\documentclass[aps,prl,twocolumn,showpacs,showkeys]{revtex4-1}
\usepackage{graphicx}
\usepackage{amssymb,amsmath}
\usepackage[latin1]{inputenc}
\usepackage{natbib}
\usepackage{epstopdf}
\usepackage{color}

\begin{document}

\title{Phase-synchronous undersampling in nonlinear spectroscopy}
\author{Lukas Bruder}
\email{lukas.bruder@physik.uni-freiburg.de}
\author{Marcel Binz}
\author{Frank Stienkemeier}
\affiliation{Institute of Physics, University of Freiburg, 79104 Freiburg, Germany}

\date{\today}


\begin{abstract}
\noindent We introduce the concept of phase-synchronous undersampling in nonlinear spectroscopy. 
The respective theory is presented and validated experimentally in a phase-modulated quantum beat experiment by sampling high phase modulation frequencies with low laser repetition rates. 
The advantage of undersampling in terms of signal quality and reduced acquisition time is demonstrated and breakdown conditions are identified. 
The presented method is particularly beneficial for experimental setups with limited signal/detection rates.
\end{abstract}



\maketitle
Ultrafast nonlinear spectroscopy in the time domain has emerged as a powerful tool to study photophysical and photochemical processes in real-time with high spectro-temporal resolution\,\cite{cho_coherent_2008, nuernberger_multidimensional_2015}. 
To meet the demand on interferometric phase stability, necessary in this type of spectroscopy, numerous stabilization methods have been developed in the past years\,\cite{fuller_experimental_2015}. 
Among these, one particularly efficient concept is the phase modulation (PM) technique,  as demonstrated in pump-probe and multidimensional spectroscopy\,\cite{tekavec_wave_2006, bruder_phase-modulated_2015, bruder_efficient_2015, tekavec_fluorescence-detected_2007, widom_solution_2013, nardin_multidimensional_2013, karki_coherent_2014, li_probing_2016}. 
This method combines acousto-optical PM with lock-in detection which drastically reduces the demands on phase stability and greatly improves the overall sensitivity. 
As an additional advantage, it can be combined with 'action' signals for detection, e.g. fluorescence\,\cite{tekavec_fluorescence-detected_2007, widom_solution_2013, bruder_efficient_2015}, photocurrent\,\cite{nardin_multidimensional_2013, karki_coherent_2014, li_probing_2016} and mass-resolved photoion detection\,\cite{bruder_phase-modulated_2015}, which improves selectivity and facilitates studies at low particle densities. 

An integral part of the PM technique is the lock-in detection which relays on modulating a signal with a well-defined frequency, followed by demodulation with a lock-in amplifier (LIA). 
This scheme works most efficient with high modulation frequencies. 
However, due to the Nyquist theorem, the modulation frequency is generally limited by the sampling rate of the experimental apparatus. 
In spectroscopy, limiting factors typically are the signal count rate, the speed of detectors or the laser repetition rate. 
This constrains the combination of lock-in detection with CCD camera based detection schemes (e.g. dispersed fluorescence\,\cite{augulis_two-dimensional_2011}, velocity map imaging detection\,\cite{eppink_velocity_1997}) and, more severely, limits the technique to high repetition rate laser systems ($\omega_\mathrm{rep} > 100$\,kHz\,\cite{tekavec_wave_2006, tekavec_fluorescence-detected_2007}). 
Yet, many applications still require operation at low sampling rates ($\omega_\mathrm{sa} < 5$\,kHz). 
In particular, in the XUV spectral range, where the PM technique shows great promise in facilitating nonlinear spectroscopy\,\cite{bruder_phase-modulated_2017}, laser sources operate mostly at 0.01-1\,kHz. 

A common strategy to overcome the Nyquist limit is undersampling, as proposed in some lock-in detection schemes, e.g. in lock-in thermography\,\cite{breitenstein_lock-thermography:_2003}, electrical impedance spectroscopy\,\cite{hartov_multichannel_2000}, in combination with random sampling\,\cite{sonnaillon_high-frequency_2008} and in case of very low signal count rates\,\cite{fink_translational_2006, phelps_single-molecule_2013}. 
In this letter, we add to this work by introducing phase-synchronous undersampling (PSU) to the PM technique. 
We identify optimum undersampling conditions and show explicitly how PSU improves the signal quality in nonlinear spectroscopy. 
In particular, we demonstrate undersampling by more than two orders of magnitude without loss of performance in signal recovery, hence, providing the opportunity to use the PM technique with sampling rates down to the 10\,Hz regime. 
Our work makes the PM approach thus accessible to a much wider range of experimental setups. 

\begin{figure}[b]
\centering
	\includegraphics[width=\linewidth]{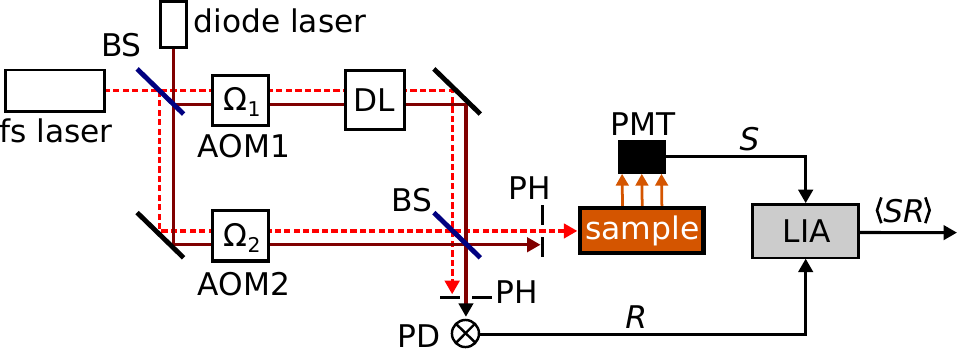}
\caption{Optical PM setup. The pump-probe signal $S$ is modulated using two AOMs driven at the frequency difference $\Omega=\Omega_2-\Omega_1$. A cw reference signal $R$ is generated with a diode laser and detected with a photo diode (PD). Pinholes (PHs) block the diode/fs laser at either of the beam splitter's (BS) exit port, respectively. Signal demodulation is done with a LIA, referenced to the cw waveform $R$. The pump-probe delay is controlled with a motorized delay line (DL). }
\label{fig:OptSetup}
\end{figure}
To demonstrate and characterize the PSU concept, we performed quantum beat measurements utilizing the PM technique in pump-probe configuration (Fig.\,\ref{fig:OptSetup}). 
An amplified femtosecond (fs) laser system of variable repetition rate ($\omega_\mathrm{rep} \leq 5$\,kHz) was used, tuned to $\lambda = 795$\,nm to drive the D1 transition in a low density rubidium vapor. 
The vapor was contained in a spectroscopy cell of which the fluorescence was detected with a photo multiplier tube (PMT). 
The PM technique and the experimental setup is described in detail elsewhere\,\cite{tekavec_wave_2006, bruder_efficient_2015}. 

Briefly, a collinear pump-probe pulse sequence is generated in a Mach-Zehnder interferometer. 
Thereby, a signal modulation ($\Omega$) is introduced with two acousto-optical modulators (AOMs) driven phase-locked at the frequency difference $\Omega$. 
The demodulation of the signal $S$ is done with a digital LIA, Stanford Research Systems, model SR 810. 
To this end, a reference waveform $R$ is constructed from a continuous wave (cw) laser ($\lambda=780\,$nm) superimposed with the optical path with a slight vertical offset (Fig.\,\ref{fig:OptSetup}), as also done in Ref.\,\cite{nardin_multidimensional_2013}. 
Throughout the data run, the pump-probe delay $\tau$ is incremented in discrete steps $\Delta \tau$ and the demodulated signal is recorded as a function of $\tau$. 
This yields the temporal evolution of electronic coherences induced in the system and a Fourier transform provides the system's absorption spectrum. 

\begin{figure}[t]
\centering
	\includegraphics[width=\linewidth]{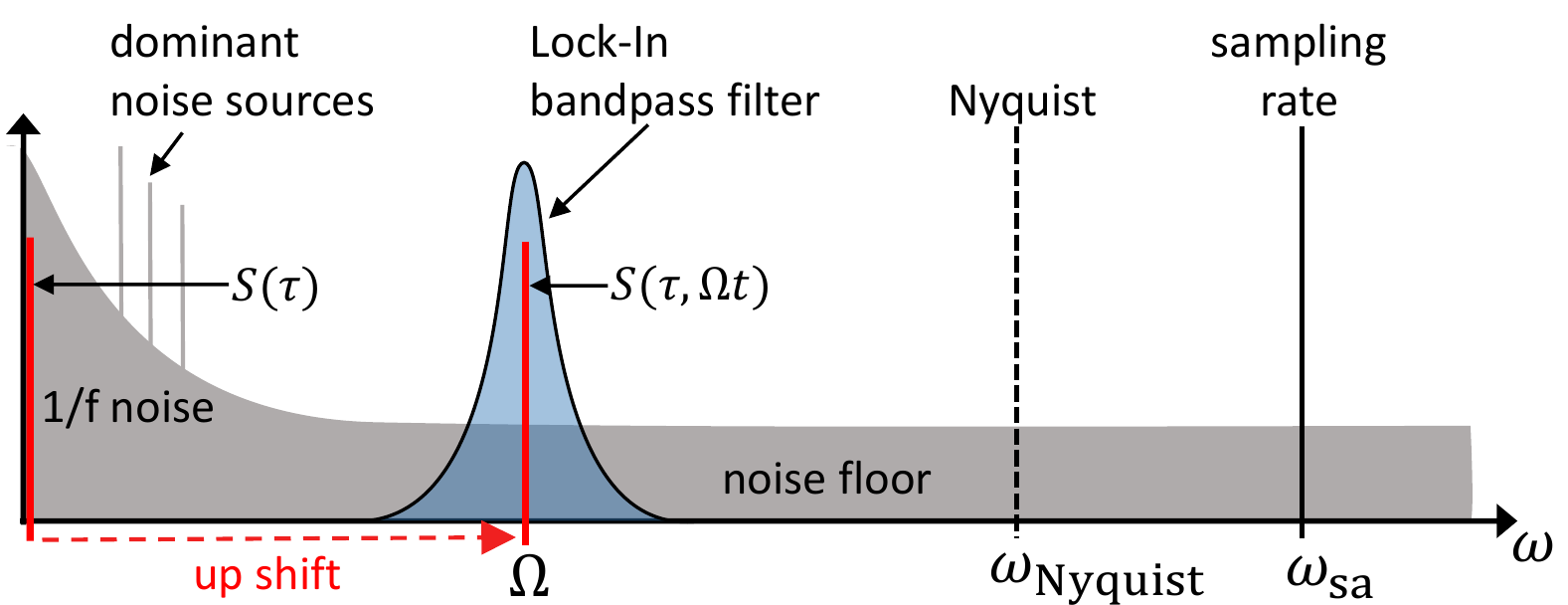}
\caption{Signal-to-noise advantage by up-shifting $S$ from low frequencies to $\Omega$ where the noise floor is much smaller and $S$ is not dominated by lab noise.}
\label{fig:SNR_Adv}
\end{figure}

As an important feature, the phase-synchronous detection of $S$ and $R$ leads to cancellation of phase jitter $\delta \phi_S$. 
Thus, yielding passive phase stabilization while the desired information is deduced from the relative phase shift of $S$ and $R$, denoted $\phi_{SR}(\tau)$. 
As a second signal-to-noise advantage, the imprinted modulation $\Omega$ shifts $S$ from the low frequency spectrum, which is often dominated by lab noise, to a spectral region with less noise, where filtering is much more efficient (Fig.\,\ref{fig:SNR_Adv}). 
For this purpose, sufficiently large modulation frequencies ($\Omega \gtrsim 1$\,kHz) are highly desirable. 
However, if the sampling rate is small, i.e. $\omega_\mathrm{sa} \leq 2\Omega$, the modulation is effectively shifted to its aliased frequency, that is 
\begin{equation}\label{eq:alias}
	\Omega_a = \mathrm{min}|n\omega_\mathrm{sa}-\Omega|\leq 0.5 \omega_\mathrm{sa},\,n \in \mathbb{Z}\, , 
\end{equation}
and the signal-to-noise advantage of using large $\Omega$ would vanish. 
In our PSU scheme, this case is circumvented due to the cw reference waveform used for the demodulation. 

\begin{figure}[t]
\centering
	\includegraphics[width=\linewidth]{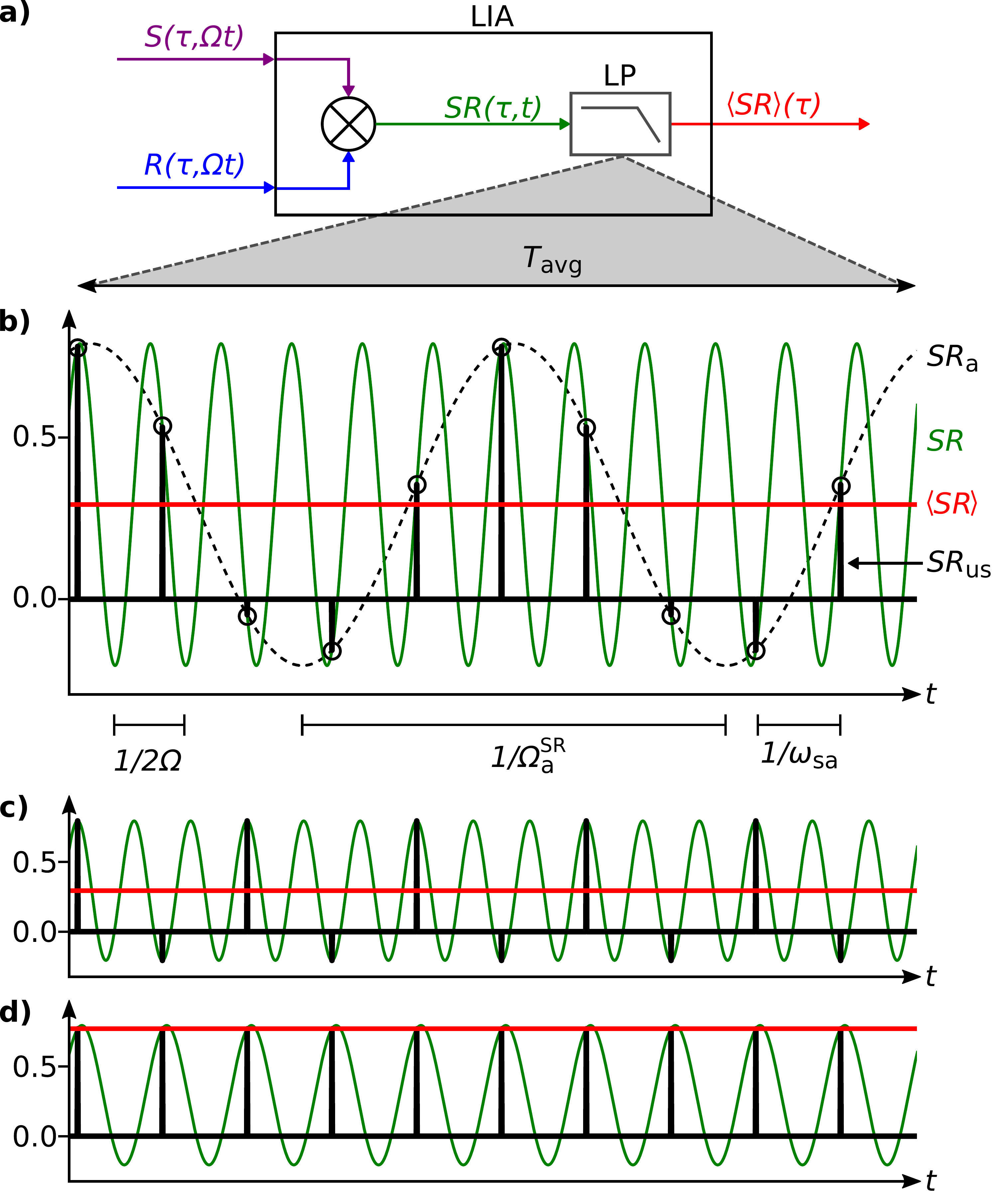}
\caption{a) Schematic description of the lock-in amplification scheme. LP denotes low-pass filter. 
b) Representation of the product signal $SR$, as constructed in the LIA. Green: fully sampled signal ($SR$), black: undersampled signal ($SR_\mathrm{us}$), dashed black: resulting aliased waveform ($SR_\mathrm{a}$), and red: mean value of $SR_\mathrm{us}$ scaled according to the sampling duty cycle ($\langle SR \rangle$). 
Undersampling of $SR$ yields the correct average value/output signal, if the averaging interval $T_\mathrm{avg}$ is sufficiently large. 
c) Optimum undersampling case. d) Breakdown case: the output value $\langle SR \rangle$ strongly depends on the relative phase between the signal modulation and the sampling points.}
\label{fig:LIAscheme}
\end{figure}

For a more detailed discussion of the PSU concept, it is sufficient to consider the excitation of a model two-level system and exemplary examine the demodulation of the in-phase signal component. 
In this case, the relevant signals $S$ and $R$ are given by\,\cite{tekavec_wave_2006} 
\begin{eqnarray}
	S(t,\tau) &=& A(\tau) \cos\left[\Omega t + \phi_S(\tau)\right] \\
	R(t,\tau) &=& \cos\left[\Omega t + \phi_R(\tau)\right] ,
\end{eqnarray}
where $A(\tau)$ and $\phi_S(\tau)$ reflect the system properties and $\phi_R(\tau)$ denotes the pump-probe dependent phase function of the reference. 
In the standard lock-in algorithm, $S$ and $R$ are multiplied and subsequently low-pass filtered (Fig.\,\ref{fig:LIAscheme}a). 
The multiplication of the two waveforms yields a signal $SR$ that consists of a sum- and difference-frequency component:
\begin{equation}
	SR (t,\tau) \propto \cos \left[ 2\Omega t + \phi_{S}(\tau)+\phi_R(\tau) \right] 
	+ \cos \left[ \phi_{SR}(\tau) \right],
\end{equation}
see also Fig.\,\ref{fig:LIAscheme}b. 
While the modulation vanishes in the difference-frequency component, the sum-frequency term exhibits a $2\Omega$-modulation. 
The subsequent RC-type low pass filter can be described by a moving average over a given time interval, denoted $T_\mathrm{avg}$\,\cite{sonnaillon_high-frequency_2008}. 
This removes the sum frequency component and thus the residual demodulated signal is 
\begin{equation}
	\langle SR \rangle(\tau) = 0.5 A(\tau)\cos \left[ \phi_{SR}(\tau) \right] .
\end{equation}
Note, that $\langle SR \rangle$ contains the desired information, that is $A(\tau)$ and $\phi_{SR}(\tau)$, while correlated phase jitter is removed ($\delta \phi_S (t) - \delta \phi_R (t) \approx 0$). 

Now, if operating in the undersampling regime, $S$ would appear at the respective aliased frequency ($\Omega_S = \Omega_a$) and since $R$ exhibits a different frequency ($\Omega_R = \Omega$), one would intuitively expect that the LIA blocks $S$ with high extinction ratio. 
However, in the lock-in amplification process, undersampling of $S$ simply leads to an undersampled product waveform of $SR$ (Fig.\,\ref{fig:LIAscheme}b). 
The resulting aliased waveform $SR_a$ exhibits the correct average value of $SR$, scaled according to the sampling duty cycle. 
Hence, as long as the mean value is calculated over a sufficient number of aliased periods, i.e. $T_\mathrm{avg} \gg 1/\Omega^\mathrm{SR}_a$, the LIA will return a correct output signal $\langle SR \rangle$. 
This explains why the PSU approach works in the PM technique. 

Note, even though $S$ is undersampled, the lock-in demodulation is performed with respect to the fully sampled frequency $\Omega$. 
Whereas, in case $S$ and $R$ are both undersampled, the phase-locked loop of the LIA will synthesize a reference waveform oscillating with $\Omega_a \ll \Omega$. 
Hence, the demodulation will be performed at the aliased frequency, which is of disadvantage in the signal recovery process (cf.\,Fig.\ref{fig:SNR_Adv}). 
Therefore, using a cw reference signal is essential. 

From the presented model, one may derive the optimum and breakdown conditions for undersampling. 
The PSU scheme works best if $\Omega^\mathrm{SR}_a$ is maximized and breaks down if $\Omega^\mathrm{SR}_a \approx 0$.
Since $SR$ is modulated at $2\Omega$, its undersampled frequency follows from Eq.\,\ref{eq:alias} under consideration of a factor of 2, which yields two extreme cases: 
%
%
\begin{align}
	(i) &: 4\Omega = (2n+1)\omega_\mathrm{sa}  & \rightarrow & \,\, \Omega^\mathrm{SR}_a = 0.5\omega_\mathrm{sa} \\
	(ii) &: 2\Omega = n\omega_\mathrm{sa}  & \rightarrow & \,\, \Omega^\mathrm{SR}_a = 0 ,
\end{align}
$n \in \mathbb{Z}$. 
(i) describes the optimum undersampling condition, where $\Omega^\mathrm{SR}_a$ reaches its maximum value and the LIA will recover the original signal correctly for minimal averaging times $T_\mathrm{avg}$ (Fig.\,\ref{fig:LIAscheme}c). 
(ii) is the breakdown case. 
Here, each sampling point of $SR$ will be at the same phase position, leading to an output signal $\langle SR \rangle$ that strongly depends on the phase offset between the imprinted phase modulation and the sampling points (Fig.\,\ref{fig:LIAscheme}d). 

The discussed model for PSU has been systematically investigated in our experimental setup. 
To this end, we performed pump-probe scans of 0-10\,ps, Fourier transformed the time-domain traces and evaluated their signal-to-noise ratio (SNR), defined as the ratio of the peak amplitude (D1 line) divided by the noise floor. 
The noise floor was deduced by calculating the mean value of the background signal in the Fourier spectrum and adding three times its standard deviation.
Error bars were estimated from the scattering of a single data point exemplary measured for ten consecutive times. 
While all other parameters were kept fixed in the experiment, the influence of the parameters $\Omega$, $\omega_\mathrm{sa}$ and $T_\mathrm{avg}$ was investigated. 
In our setup, signal rates were sufficiently high and a fast photo detector was used, therefore $\omega_\mathrm{sa}$ is defined by the laser repetition rate $\omega_\mathrm{rep}$ which was varied by pulse picking. 
$T_\mathrm{avg}$ was varied by changing the lock-in time constant $T_\mathrm{LI}$ and keeping the filter roll-off at 6\,dB. 

\begin{figure}[t]
\centering
	\includegraphics[width=\linewidth]{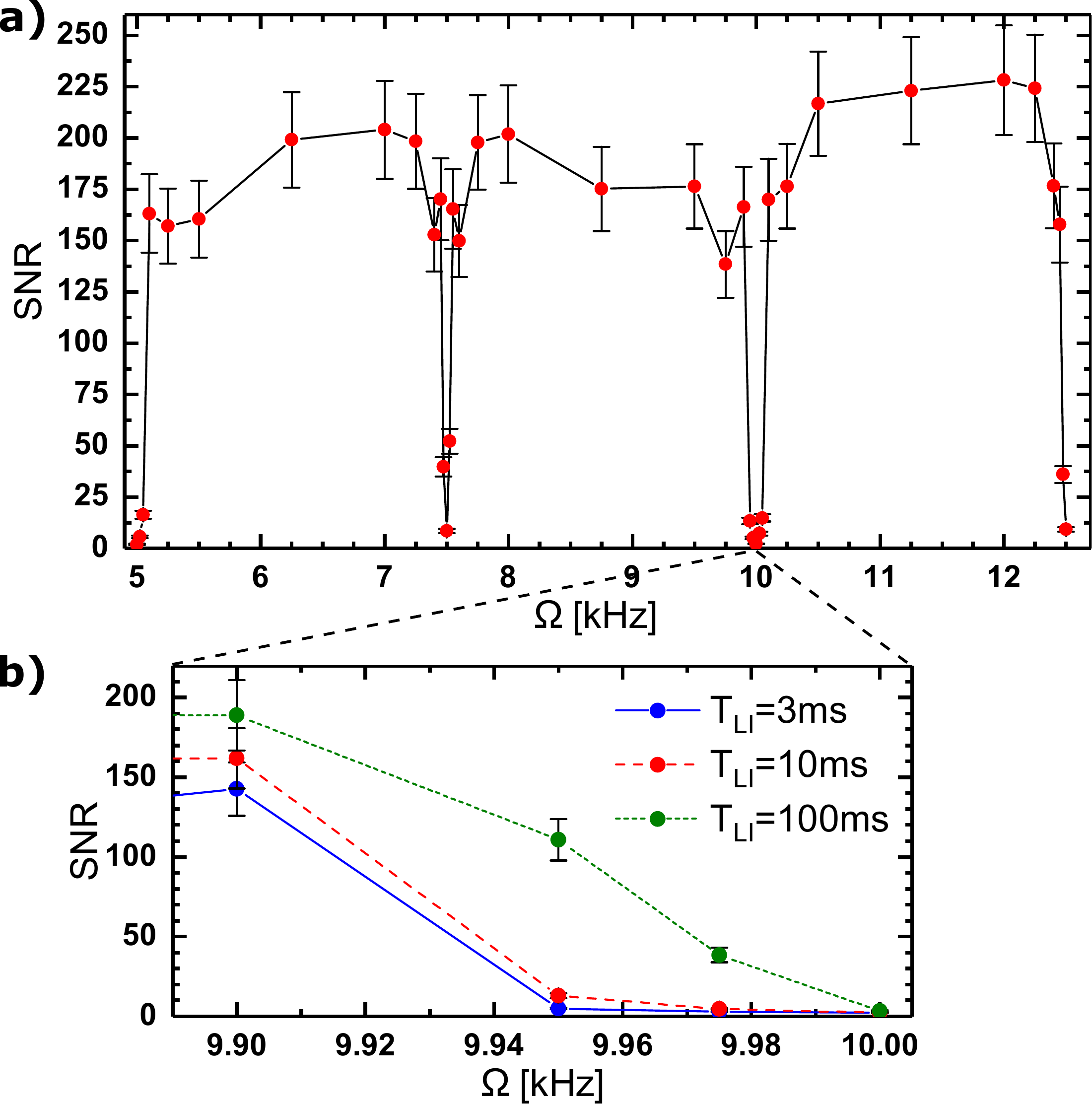}
\caption{a) SNR as a function of the modulation frequency $\Omega$, recorded for a fixed sampling frequency of $\omega_\mathrm{sa}=5$\,kHz. 
b) Comparison of different averaging intervals $T_\mathrm{avg}$, varied via the lock-in filter time $T_\mathrm{LI}$. }
\label{fig:Cases}
\end{figure}
At first, the undersampling cases (i) and (ii) are characterized. 
Fig.\,\ref{fig:Cases}a shows a scan of the modulation frequency $\Omega$ for a fixed laser repetition rate of $\omega_\mathrm{rep}=5$\,kHz and a lock-in time constant of $T_\mathrm{LI}=10$\,ms. 
As expected, for condition (ii), the SNR drastically decreases. 
This gives rise to 'forbidden' sampling frequencies at multiplies of the Nyquist frequency, in accordance to previous reports\,\cite{krapez_compared_1998}. 
In between, the SNR forms a plateau centered at condition (i). 
The fairly large scattering of SNR values in the plateau region stems from fluctuating experimental conditions. 
These induce significant variations in the signal, due to the relatively short filter time constant used in this measurement series. 

Fig.\,\ref{fig:Cases}b shows the comparison of different filter times for a zoom on a forbidden frequency. 
With increasing filter time, i.e. increasing averaging time interval $T_\mathrm{avg}$, the slope becomes steeper. 
This is in accordance with the PSU model, as it predicts, that with increasing $T_\mathrm{avg}$, smaller aliased frequencies $\Omega^\mathrm{SR}_a$ will still lead to a correct demodulated output signal (cf. Fig.\,\ref{fig:LIAscheme}b-d). 
Note, that in general, lower sampling rates lead to more narrow SNR plateaus, making it more difficult to find ideal undersampling conditions. 
This can be diminished to some extend using larger lock-in time constants (Fig.\,\ref{fig:Cases}b). 
As such, we were able to work also at $\omega_\mathrm{rep}=50$\,Hz, $T_\mathrm{LI}=1$\,s without significant loss of performance. 

In overall, with this measurement series we have confirmed the theory of PSU. 
For the discussed extreme cases (i) and (ii), the experiment behaves as predicted and the qualitatively correct dependency on the average time interval/lock-in filter time constant has been shown. 

In a second study, we focused on the actual signal-to-noise advantage of PSU. 
To this end, we compared for two sampling rates ($\omega_\mathrm{rep}=0.5$ and 5\,kHz) a scan of $\Omega$ from 50\,Hz to 41.25\,kHz. 
The ratio $\Omega / \omega_\mathrm{rep}$ was chosen such, that condition (i) was always met. 
Furthermore, we used $T_\mathrm{LI}=100$\,ms, i.e. $T_\mathrm{avg}=const.$ and at low modulation frequencies ($\Omega \leq 200$\,Hz), the LIA was set to sync mode.  

\begin{figure}[t]
\centering
	\includegraphics[width=\linewidth]{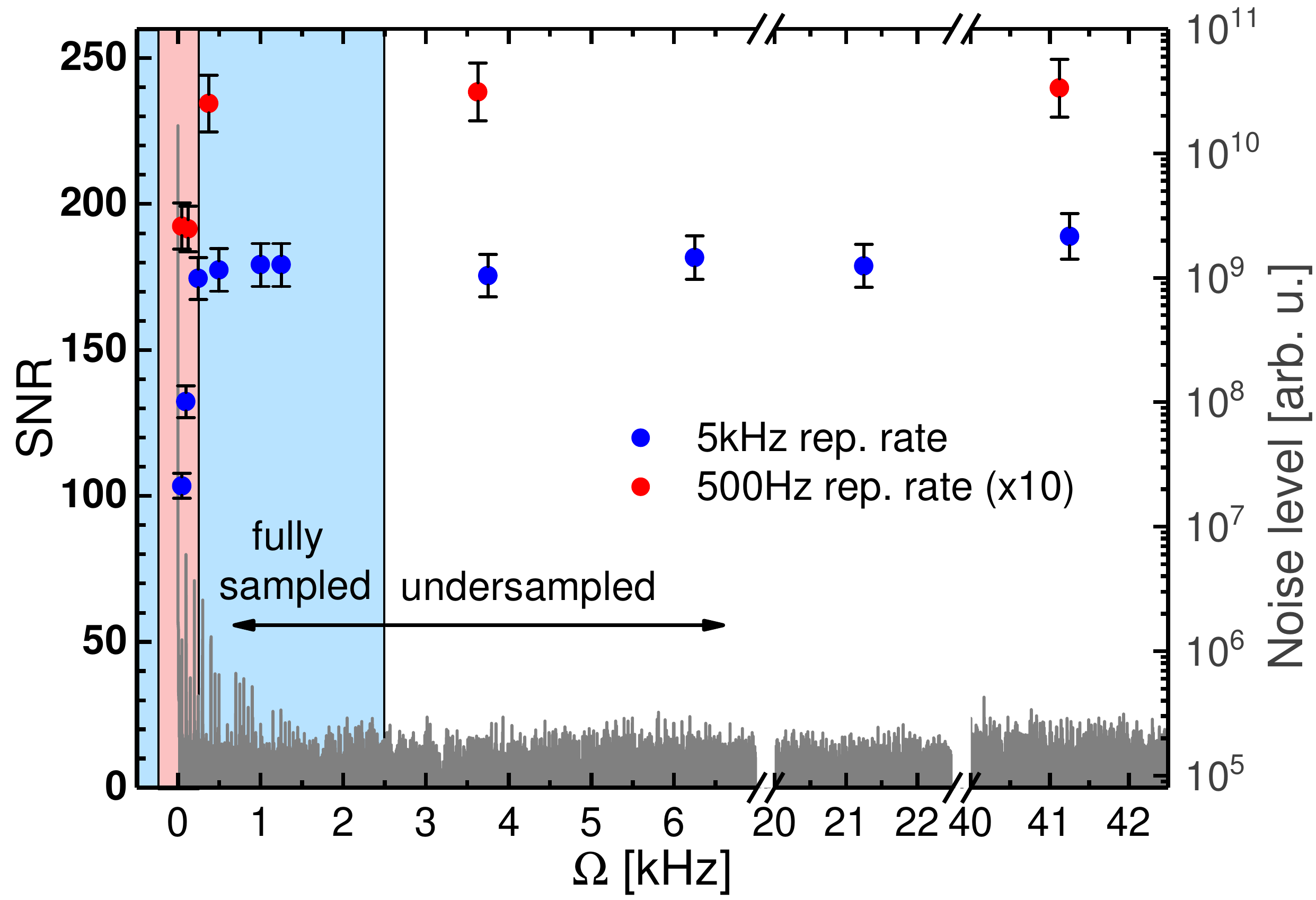}
\caption{SNR for fully and undersampled signal modulation, measured for laser repetition rates of $\omega_\mathrm{rep}=5$ and 0.5\,kHz. The latter data set was scaled by a factor of 10 for better visibility. Blue and red shaded areas indicate the fully sampled regime for the two data sets, respectively. Grey: noise spectrum of the lab, as picked up in the experiment.}
\label{fig:PSU}
\end{figure}

Fig.\,\ref{fig:PSU} shows the result. 
At low modulation frequencies ($\Omega \lesssim 250$\,Hz) the SNR decreases whereas at high frequencies the data quality remains constant, even for very high undersampling factors (demonstrated up to a factor of 166). 
Thus, the PSU concept allows choosing very high modulation frequencies without loss of performance. 
Interestingly, for $\omega_\mathrm{rep}=5$\,kHz (blue), the optimum SNR value can be reached within the fully sampled regime, however, this is not the case for the $\omega_\mathrm{rep}=0.5$\,kHz data (red). 
Hence, in case of low sampling/laser repetition rates, undersampling clearly improves the signal quality. 

The reduction of the SNR at low frequencies has two reasons. 
One is due to the forbidden frequency at 0\,Hz. 
As discussed, larger lock-in time constants would attenuate this effect. 
However, this comes at the price of longer lock-in settling times and hence, increasing acquisition times. 
For instance, in order to reach in the fully sampled regime the same SNR as obtained in the undersampled case (referring to red data, Fig\,\ref{fig:PSU}), the acquisition time would increase by roughly an order of magnitude. 

The other and more important reason, however, is the increasing lab noise at low frequencies. 
To visualize this, the lab noise spectrum was deduced from a Fourier analysis of the reference signal $R$ at a fixed pump-probe delay (Fig.\,\ref{fig:PSU}). 
Note, the noise spectrum is plotted in a log scale while the SNR is given in a linear scale. 
In nonlinear spectroscopy, low frequency noise sources coming, for instance, from the power line, harmonics thereof, ground loops or mechanical vibrations, can easily dominate the signal at low frequencies. 
In this regard, shifting the signal to higher frequencies clearly improves the data quality, as confirmed by our measurements. 

For high modulation frequencies, we find, that the signal amplitude is damped due to the bandwidth limit of the detection electronics (not shown). 
This is surprising, since the electronics should only sense the aliased frequency $\Omega_a < 0.5\omega_\mathrm{rep}$ of the signal $S$. 
We explain this by phase jitter in the signal modulation caused by phase fluctuations in the optical interferometer. 
This phase/timing jitter relative to the laser shots can lead to high frequency signal components which are damped according to the bandwidth limitation of detection electronics, thus reducing the overall signal. 

We point out, that in the applied lock-in detection, the signal quality depends in a non-trivial way on two factors, that is the general signal recovery capability of the lock-in algorithm and the accompanied phase stabilization effect. 
As such, results from undersampling strategies proposed in other fields\,\cite{hartov_multichannel_2000, breitenstein_lock-thermography:_2003} are not directly transferable. 
Our work thus introduces a new detection concept in phase-modulated nonlinear spectroscopy. 
In the presented study, the sampling rate was determined by the laser repetition rate, however, the PSU scheme also works if the sampling rate is limited by other factors, e.g. detector speeds or low signal count rates. 

In conclusion, phase synchronous detection with a LIA has particular advantages in nonlinear spectroscopy as it greatly improves sensitivity and at the same time introduces passive phase stabilization. 
In this context, we have introduced the concept of PSU for nonlinear spectroscopy which facilitates efficient lock-in detection at high modulation frequencies ($\sim $kHz) while using sampling rates down to the 10\,Hz regime. 
We have identified and experimentally verified the optimum and breakdown conditions in this approach and demonstrated that PSU improves the signal quality and increases acquisition speed, especially if the experimental apparatus is limited to low sampling rates ($\leq 500$\,Hz). 
Limiting factors are often the laser repetition rate, slow detectors or low signal count rates, hence, our concept makes the PM technique feasible for much more experimental setups. 
In particular, the application in nonlinear XUV spectroscopy may have great potential\,\cite{bruder_phase-modulated_2017}. 

\noindent \textbf{Funding.} Bundesministerium f\"ur Bildung und Forschung (BMBF), project 05K16VFB; Deutsche Forschungsgemeinschaft (DFG), program IRTG 2079

%

\end{document}